\documentclass[12pt]{iopart}
\pdfoutput=1
\usepackage{graphicx}
\usepackage[caption=false]{subfig}
\usepackage{color}
\usepackage{float}

\graphicspath{{./Figures/}}

\begin{document}

\title[]{Testing macroecological theories in cryptocurrency market:
neutral models can not describe diversity patterns and their variation
}
\author{Edgardo Brigatti$^{\star}$  and  Estevan Augusto Amazonas Mendes}

\address{ Instituto de F\'{\i}sica, Universidade Federal do Rio de Janeiro,
Av. Athos da Silveira Ramos, 149,
Cidade Universit\'aria, 21941-972, Rio de Janeiro, RJ, Brazil}
\address{$^{\star}$ e-mail address: edgardo@if.ufrj.br}

\maketitle

\begin{abstract}
We develop an analysis of the cryptocurrency market borrowing methods and concepts from ecology.
This approach 
makes it possible to identify 
specific diversity patterns and their variation, in close 
analogy with ecological systems, and to characterize 
the cryptocurrency market in an effective 
way.
At the same time, it shows how non-biological systems can have an important role
in contrasting different ecological theories and in testing the use of neutral models. 
The study 
of the 
cryptocurrencies abundance distribution 
and the evolution of the community
structure strongly indicates that these statistical patterns are not consistent with neutrality.
In particular, the necessity to increase the temporal change in community composition 
when the number of cryptocurrencies grows, suggests that their interactions 
are not necessarily weak.  
The analysis of the intraspecific and interspecific interdependency supports this fact and
demonstrates the presence of a market sector influenced by mutualistic relations.
These latest findings challenge
the hypothesis of weakly interacting symmetric species,
the postulate at the heart of neutral models. 

\end{abstract}



\section{Introduction}

Since the appearance of Bitcoin, the first peer-to-peer digital currency, 
introduced by a paper 
authored under the pseudonym of Satoshi 
Nakamoto in 2008 \cite{Nakamoto2008}, 
the cryptocurrency 
market has known a spectacular growth in terms of capitalization 
and the number of 
different cryptocurrencies. 
This success arouses economists' interest,
engaged in exploring if these digital currencies could 
perform all the three functions of money (medium of exchange, unit 
of account, and store of value) \cite{Ammous} 
and worried about the effects that this trend could produce on 
traditional monetary or governmental authorities. 

As most cryptocurrencies are used more like a portfolio speculative asset than as a currency,
they attracted the attention of traders and institutional investors.
This trend 
has been reflected in academic studies, that generally have focused
on Bitcoin and its price dynamics \cite{finance1,finance2,Mio}. 
More recently, some studies have 
tried to describe the overall market \cite{polacco}, looking at the dynamics  
of all the cryptocurrencies actively traded 
\cite{baronca,Wu}.
Among these last works, El Bahrawy {\it et al.} \cite{baronca}
introduced an interesting perspective
looking at the cryptocurrencies market as an ecological system.
Based on this frame of mind, they described several typical distributions. 
The use of this parallelism is not new and follows a stream of works
where non-biological \cite{Gaston,Blonder}, synthetic systems \cite{Linux,Valverde,Bedau}  
or artificial-life type simulations \cite{Chow} have been described in terms of ecological systems and 
used for testing ecological and evolutive  theories.
In this work, we will focus on these aspects, developing an in-depth analysis
of the cryptocurrencies market either 
for describing some of its fundamental aspects  
or for comparing and evaluating 
ecological models.\\ 


In the last two decades, two contrasting ecological theories generated
an important debate around the principal mechanisms
that shape ecological communities. 
Niche theories \cite{Grinnell,MacArthur} 
state that the most relevant factors that determine
communities are defined by the selection produced by the interactions 
among the individuals, the species, and the environment. 
In contrast, neutral theories \cite{hubbell01,Volkov03} consider that the dominant factors are 
the stochastic processes present in populations dynamics,  
which determine their random drift.



In its more successful implementation, the neutral theory of biodiversity
models the organisms of a community with identical per-capita
birth, death, immigration and speciation rates \cite{hubbell01}. 
There are no differences among the species, which are considered demographically and 
ecologically identical. These hypotheses imply, from a theoretical perspective, the
assumption of functional equivalence: on the same trophic level
species are characterized by identical rates of vital events. 
From a model 
perspective, they lead to 
the symmetry postulate: the dynamics of the 
community is not influenced by interchanging the species labels of individuals \cite{Borile}.

In this work we will consider only neutral models which describe
populations at the level of individuals, not of species,
allowing the direct estimation of species abundances.  
Among these models, we consider very general 
Markovian models, usually described through a master equation \cite{Volkov03}
or a Langevin equation \cite{Azaele06}.
These approaches generate predictions at stationarity. 
For this reason, we will focus our attention on 
data that can be considered closed to stationarity.

These ecologically neutral models generate statistical neutral
distributions of different macroecological patterns.
In general, to test these theories with empirical data
it is determined, using a statistical selection, 
if such distributions are compatible with the empirical ones.
Strictly speaking, this statistical neutrality (the adherence between the model and data 
distribution) is a necessary but not sufficient condition for claiming neutrality \cite{Fisher14}.
Even if this is the most pragmatic way for testing neutrality, the analysis of the 
form of interactions and if species satisfy the assumption of functional equivalence 
could produce more enlightening and definitive results.

Based on these considerations, 
the principal aim of our study 
will be the analysis 
and characterization of different macroecological patterns converted to the specific case of
the cryptocurrency market.
Furthermore, we will focus our attention on the examination 
of the interdependence and correlations present among cryptocurrencies. 
In the section on Methods, we will present in detail which patterns and how these analyses 
can be carried out.\\

In brief, the construction of an analogy between the cryptocurrency market and ecological system
will better elucidate the community structure of the cryptocurrency market shedding light on the existing relationships between cryptocurrencies.
Despite its importance, the theoretical description of 
structure and interactions in markets 
is limited and new concepts and approaches are required.
Ecology is capable of introducing new and powerful ideas, 
previously debated and  tested.
In this work, we will describe specific dynamics, determine new stylized facts and 
patterns and compare them with a testable theory.
This analysis will bring new insights and comprehension of the considered market, highlighting important elements which could be useful for assisting in hedging risks.
The natural limitation of an analogy-based approach could
be  overcome in the future by introducing specific features 
for correcting inaccuracies and building new models.

\section{Data}

We collected the cryptocurrencies data from the website Coin Market Cap \cite{CoinMarketCap2020},
which extracts from the 
exchange market platforms the price expressed in US dollars (exchange rate), the volume of trading in the preceding 24 h and the market capitalization of the different cryptocurrencies.
Market capitalization is the product of the price for the circulating supply, which 
can not account for dormant or destroyed coins.
Traded cryptocurrencies can disappear from the website list to reappear later on. In fact,
capitalization of cryptocurrencies not traded in the 6 h preceding the weekly report is not included in
the dataset and cryptocurrencies inactive for 7 days are not included in the list.
For this reason, we filled these lacunae by introducing the average between the 
values available at the extremes of each gap. Finally, we cleared the dataset by 
correcting some mismatches or typos present in the names or symbols of the cryptocurrencies.

We considered weekly data from 28 April 2013 to 2 February 2020, which correspond to a series of 354 time steps. 
The dataset contains a total of 3588 cryptocurrencies.


\section{Methods}

To construct the parallelism between ecological systems and the cryptocurrency
market we consider that each cryptocurrency represents a species and its capitalization
its abundance. In this way, the wealth invested in a cryptocurrency replaces 
the population size of a species.

The first macroecological pattern that we analyze is the Species Abundance Distribution (SAD), 
a classical biodiversity descriptor that characterizes static features of  
ecosystem diversity.  This distribution represents 
the probability that a species presents a given population size. 
By using a very general stochastic Markovian model for neutral ecological communities,
Volkov {\it et al.} showed that, at stationarity, the SAD 
follows an analytic zero-sum multinomial distribution 
for local communities and, for the metacommunity, the celebrated Fisher's 
distribution  \cite{Volkov03,McGill06}, as already predicted by Hubbell's neutral model \cite{hubbell01}.
Note that, as the model is   
nonspatial
the term metacommunity represents 
a single, permanent large community where migration is absent. 
In contrast, local communities present  dynamics defined by immigration from 
a permanent source 
of species (the metacommunity).
In this sense, our data must be considered as collected from the metacommunity
and, based on the results of  Volkov {\it et al.}, 
they should be described by 
a Fisher's distribution: 
$p(x)\propto e^{-cx}/x$ \cite{Fisher43}.
A classical alternative 
to the distribution produced by the neutral theory 
is the log-normal one: 
$p(x)=\frac{1}{\sqrt{2\pi \sigma^2}x} e^{-(\log{x}-\mu)^2/2\sigma^2}$,
which is equivalent to a normal distribution if the variable $log(x)$ is chosen.
This distribution has a long tradition of use in the ecological literature \cite{Preston48} and it
is a reasonable and parsimonious null hypothesis for the SAD \cite{McGill_Nature03}, 
which can be obtained from pure statistical non-biological  arguments, based on the central limit theorem \cite{May75},   
or  can be generated by niche or demographic differences among species in population models \cite{Connolly_PNAS14}.  
To sum up, a description of the SAD with a Fisher's distribution
would support the idea of  Neutral dynamics. In contrast, 
a log-normal 
distribution would suggest the presence of other types of 
population dynamics or pure statistical mechanisms.

The second macroecological pattern is the Species Population Relation (SPR).
It describes the scaling of the number of species $N$ with the total number of individuals $x$.
When the different 
species of a total population $x$ follow the Fisher's 
distribution,
the expected number of species for a given population $x$ is given by:
$N(x)= \alpha \log{(1+\frac{x}{\alpha})}$ \cite{Fisher43}.  
Ecological studies  usually measure the relationship between 
the number of species and the area sampled, which can be easily obtained assuming 
a linear relation between $x$ and the area.
An alternative characterization of this relation commonly found in the literature is based on empirical curves showing a power-law behavior, with  exponents presenting typical values between 0 and 1 \cite{Azaele16}. \\ 



So far we have focused on static macroecological patterns
which can be related to results generated by
neutral theories at the steady-state.
However, stationarity is just an approximation, and not necessarily a good one,
for systems characterized by a state of flux in species, abundance and composition.
For this reason, we look at the temporal behavior of our system and characterize 
time-dependent patterns.

The intertwined history of different species along their evolution generates  
complex communities characterized by non-trivial structures.
New species continuously replace old ones producing an intricate overlap where the turnover 
can be characterized employing  the species turnover distribution (STD).
This distribution is defined as the probability that the ratio of the population of a species separated by a time interval $t$, $x(t)/x(0)$, is equal to $\lambda$.   
Under stationary conditions, Azaele {\it et al.} \cite{Azaele06} introduced a neutral model that can forecast the STD.
Within this framework,  an analytic expression for this distribution is obtained:
\begin{equation}
P_{STD}(\lambda,t)=C\frac{\lambda+1}{\lambda}\frac{(e^{t/B})^{A/2}}{1-e^{-t/B}}\bigg[\frac{sinh(t/2B)}{\lambda}\bigg]^{A+1}\bigg[\frac{4\lambda^2}{(\lambda+1)^2e^{t/B}-4\lambda}\bigg]^{A+1/2}
\label{eq:STD}
\end{equation}
where  $A$ and $B$ are 
parameters and $C$ is a renormalization constant equal to 
$\frac{2^{A-1}}{\sqrt{\pi}}\frac{\Gamma(A+1/2)}{\Gamma(A)}$.
 

The community composition of cryptocurrencies presents a coherent structure
that evolves over time. 
This structure can be quantified by analyzing how the 
capitalization of each coin changes. 
A variety of different indices has been used in ecology studies for 
tracking 
modifications in community composition through a similarity measure
\cite{Dornelas14,Magurran19}. Here, we adopt a simple and well-known one that presents 
a straightforward interpretation of its values \cite{McGill_PNAS05}.
Given the capitalization $C_i(t)$ of a given species $i$, 
we consider $S_i^t=log(C_i(t)+1)$ (1 is added 
in order to avoid the log of $0$),
where $t$ is a given month.
Note that the log-transformation is a natural approach for quantities which can be
roughly described by a log-normal distribution. 
The next step is the estimation of the Pearson's correlation of the log-transformed data:
$r_S(\tau)=Corr(S_i^{t_0},S_i^{t_0+\tau})$, where the correlation is calculated over the index $i$.
This method 
is well known  \cite{Engen02,McGill_PNAS05} 
and easily supports tests of significance: 
0 represents complete randomness and is the null hypothesis for significance tests.
To conclude, we 
characterize the mode 
of the community similarity evolution by
looking at the functional form of the temporal decay of $r_S$, which describes the transformation 
from perfectly correlated structures towards totally uncorrelated ones.

We compare the results obtained from this empirical analysis of the community composition 
with the ones generated by a neutral model introduced in \cite{Pigolotti,Azaele06},  
which is constituted  by a system of stochastic discrete differential equations.
The model describes an ecological community with a fixed number $s$ of species, where
the total number of individuals in the community is set to $N$.
In the large population limit, these constraints can be relaxed and similar results are obtained  \cite{Pigolotti}.
If $x^t_i$ represents the population of the i-th species at time $t$, 
its evolution follows this equation: 
\begin{equation}
x^{t+1}_i=N \frac{\rho x^t_i+\sigma \sqrt{x^t_i}\eta^t_i +b   }{\sum^s _{j=1} (\rho x^t_j+\sigma \sqrt{x^t_j}\eta^t_j +b)}.
\label{equ_model}
\end{equation}
The parameter $b$ controls the population size near the extinction threshold,
taking into account the total effect of immigration, extinctions and speciations.
$\rho$ and $\sigma$ are the mean value and the standard deviation of the distribution
which describes the per capita birth rate.
Based on the principle of neutrality, we assume that $\rho$ and $\sigma$ are 
the same for all individuals. Finally, $\eta^t_j$  is 
an uncorrelated Gaussian noise term, with zero mean and unit variance.
By taking the continuum limit, this stochastic discrete model can be approximated by a
Langevin and the associated Fokker-Plank equation.
Its analytical solution shows that, in the limit of large $N$, 
a Gamma distribution describes the SAD 
for this model.
Such distribution can well describe various experimental SAD data and
approaches the Fisher's distribution 
for $2b/\sigma^2<<1$ \cite{Pigolotti,Azaele06}. 
These results can be connected with the classical outcomes of the neutral theory 
as formulated by Volkov {\it et al.}  \cite{Volkov03} for a particular choice of 
the birth and death coefficients.\\


Our last analysis looks at the characterization of the interdependence among cryptocurrencies. 
This inspection will help in exploring  if in a community of cryptocurrencies the 
symmetric species postulate is well supported 
and if interactions between species can be considered weak in relation to stochastic effects,
as expected for neutral dynamics.
As capitalizations present a growing trend, we do not study directly their correlations.
In contrast, we consider the log-variation of the capitalization of a given species $i$:
$V_i(t)=ln[C_i(t)]-ln[C_i(t-1)]$.
As a dependence measure, we consider the Pearson's correlation between the synchronous 
time evolution of $V(t)$ of a pair of cryptocurrencies $i$ and $j$:
$r^{ij}_V=Corr(V_i(t),V_j(t))$.
The interdependencies among the log-variations in the capitalization
are addressed looking at the corresponding correlation matrix.

\section{Results}

\subsection{General dynamics}

\begin{figure}[h]
\begin{center}
\includegraphics[width=0.95\textwidth, angle=0]{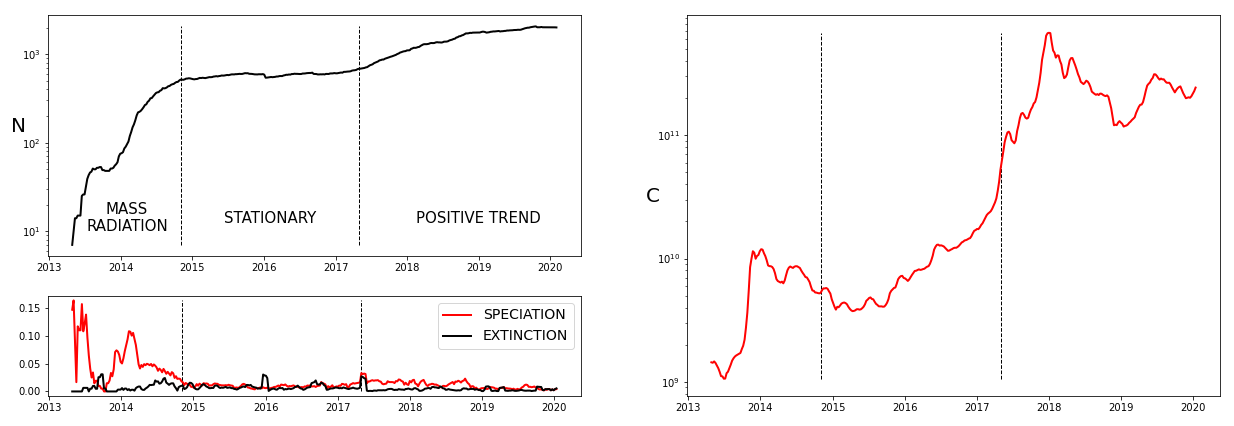}
\end{center}
\caption{\small {\it On the left:}  number of actively traded cryptocurrencies (top); speciation and extinction rates during a given week averaged over rolling windows of 4 weeks (bottom). The dashed
vertical lines represent the starting of the stationary phase (2014-11-02)  and its ending (2017-04-30).
{\it On the right:} market capitalization averaged over rolling windows of 4 weeks.}
\label{fig_rates}
\end{figure}

We consider the time series which collect the number of active cryptocurrencies at a given 
time ($N(t)$) and we evaluate the corresponding 
speciation and extinction rates (see Figure \ref{fig_rates}), measured as the number of cryptocurrencies entering (respectively, leaving) the market on a given week, normalized over the number of active cryptocurrencies at that time. 
By looking at the number of active cryptocurrencies and at the 
speciation and extinction rates, it is possible to qualitatively discriminate between 3 different regimes:
before  2014-06-01 a mass radiation phase,
characterized by a spectacular increase in the number of 
cryptocurrencies, caused by a very high number of speciation events. 
This phase is characterized by significant fluctuations. 
Between 2014-11-02 and 2017-04-30 we can highlight a stationary phase,
with comparable values of speciation and extinction rates.
Starting from 2017-05-07, a positive trend characterizes a new regime where the number
of cryptocurrencies 
grows slightly in a gradual and regular fashion. Also, this regime
presents higher values of speciation rates in relation to the extinction ones.
In general, extinction rates fluctuate around a typical value, except for some
sparse and sudden extinction bursts. In contrast,
speciation rates present different behaviors which determine the modes of 
the three considered regimes.

The total market capitalization $C(t)$ in general presents important positive trends.
In the regime where the number of active cryptocurrencies is stationary, intervals of 
exponential growth can be detected;
in the regime with important radiation of new currencies, stages 
with super-exponential behaviors can be identified.



\subsection{Species Abundance Distribution}

\begin{figure}[h]
\begin{center}
\includegraphics[width=0.9\textwidth, angle=0]{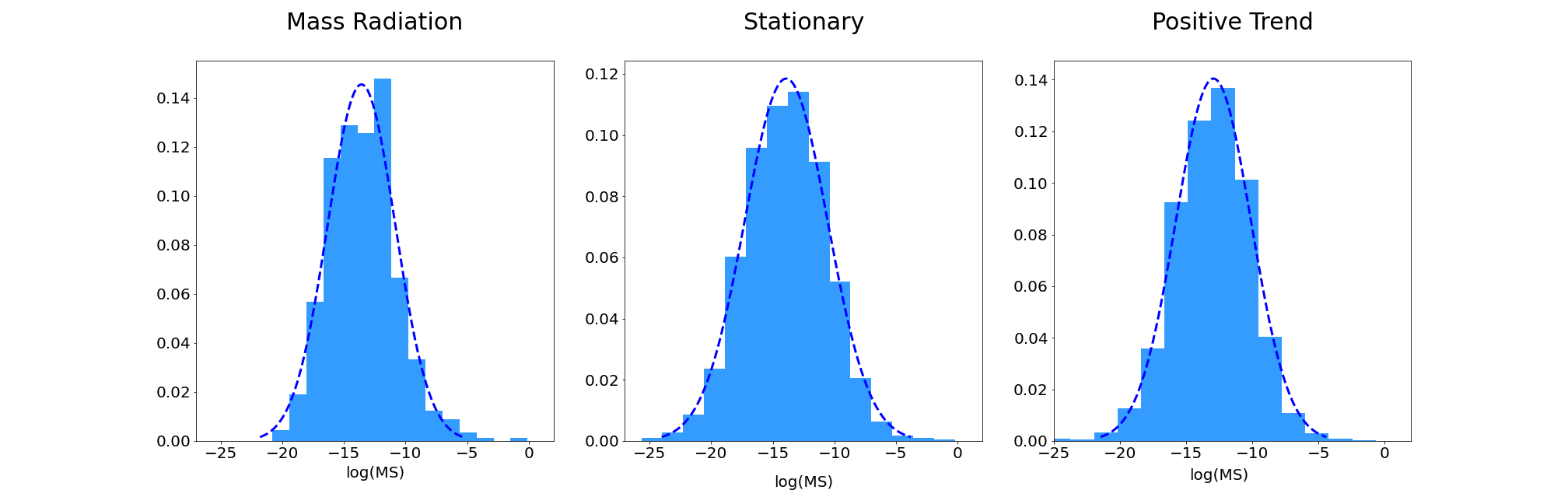}
\includegraphics[width=0.9\textwidth, angle=0]{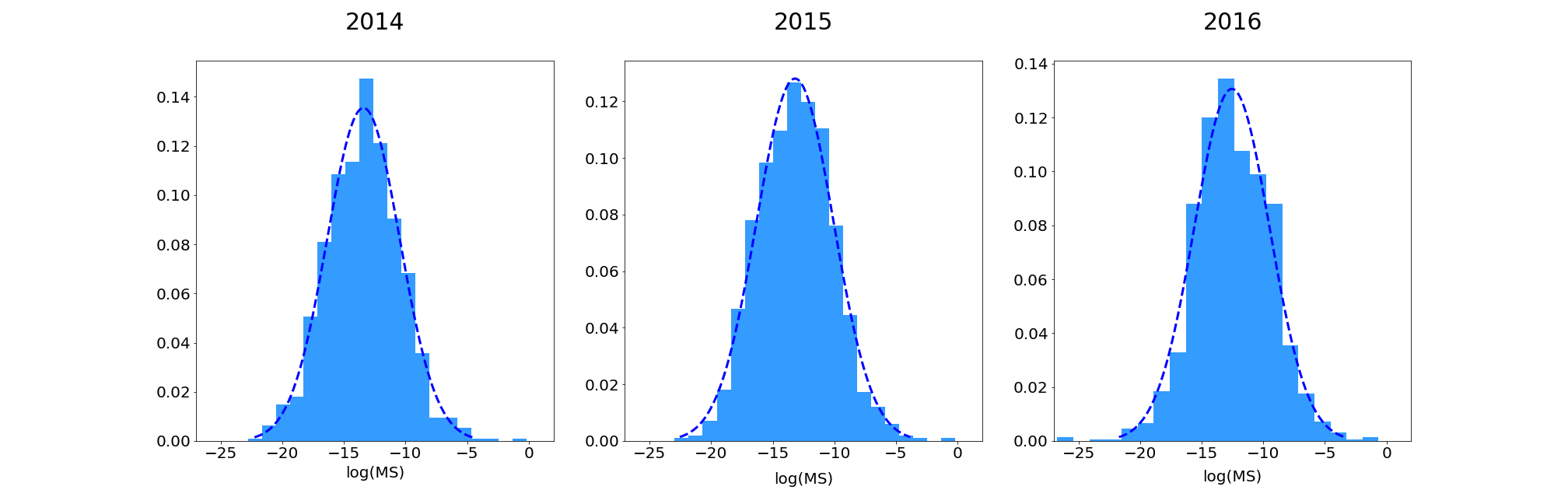}
\end{center}
\caption{\small { Species Abundance Distribution sampling the empirical data along different periods.
{\it Top}: data sampled during the periods of mass radiation, the stationary period and
the positive trend. The p-values of the  Kolmogorov-Smirnov test are 
0.584, 0.626, 0.245 for datasets presenting 652,1291 and 2825  elements. 
{\it Bottom}: data collected during the stationary period, starting in november 2011,
and sampling along one year. The p-values of the  Kolmogorov-Smirnov test are
0.996, 0.715, 0.253 for datasets presenting 840, 866 and 1170  elements.
}
}
\label{Fig_RSA}
\end{figure}

Figure \ref{Fig_RSA} shows the SADs considering different 
periods and time scales.
We present data aggregated along a year or more, 
considering all the stationary and non-stationary periods.
For each cryptocurrency, we measure the corresponding
market share $(MS)$ for overcoming the problem of 
the non-stationarity of the capitalization. 
Comparable results can be obtained for shorter time scales, corresponding
to a month or even a week.
For testing the neutral theory we must focus on the stationary period, where we can consider
the dynamics in the number of active cryptocurrencies as stationary 
and we can compare the empirical data to the theoretical predictions, which are
obtained at stationarity.
It is important to note that, in ecology, the situation of incompletely censused regions is frequent. This condition usually produces an under-sampling of rare species.
In this situation, it is common to consider to fit only the portion of the distribution which 
encompasses the most common species. 
In contrast, in our case, we can consider having access to a fully sampled
community, an assumption supported by the regularity of the shape of the left tail of the distribution,
which does not suffer from strong statistical fluctuations. 
Just by looking at the full shape of our distributions, which always present an evident internal mode,
we can firmly discard the hypothesis that Fisher's distribution can give a good description of our
dataset. In fact,  Fisher's distribution is monotonic and does not present internal modes. 
In contrast, we can 
perform a high-quality fitting using the log-normal distribution.
In particular, by 
using a Kolmogorov-Smirnov test of our observations against 
the fitted log-normal distributions,
we cannot reject the hypothesis that 
the data come from the fitted distribution since  p-values are very high in all the considered
cases (see Fig. \ref{Fig_RSA}).


\subsection{Species Population Relation}

\begin{figure}[h]
\begin{center}
\includegraphics[width=0.8\textwidth, angle=0]{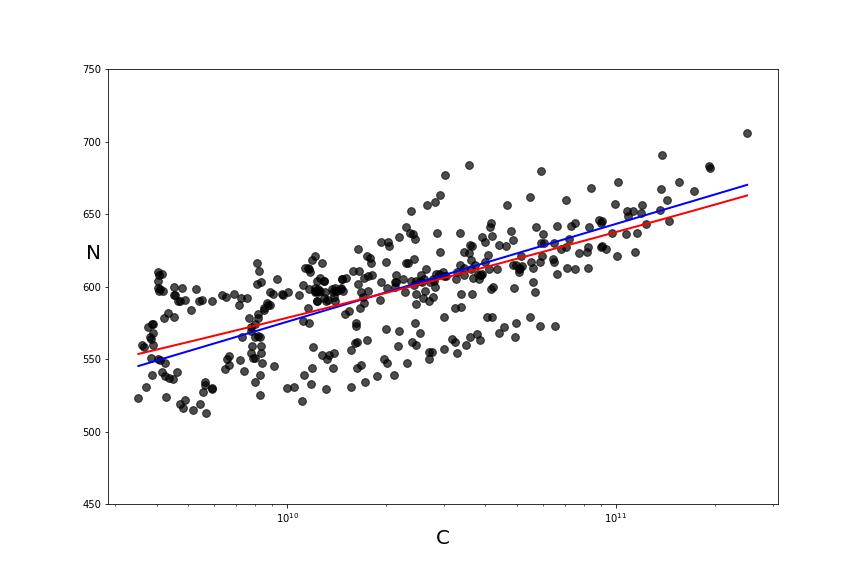}
\end{center}
\caption{\small {Log-linear plot of the capitalization versus the number of species.
The red line is the best fitting obtained from the relation: $\alpha \log{(1+\frac{x}{\alpha})}$, 
the blue one from a power law.
}
}
\label{Fig_SPR}
\end{figure}

For studying the SPR,  we analyze the stationary period and scan the corresponding dataset with different temporal intervals. We consider intervals between 1 and 10  weeks and, for each sample, we measure the  value of capitalization 
and number of species.
We plot the pairs of all these values in Figure \ref{Fig_SPR}, where we represent 
the fitting obtained by using 
the relation $\alpha \log{(1+\frac{x}{\alpha})}$ compared with a power-law.
The dependence of $N$ on $C$ is very weak and it is not possible to affirm if 
the logarithmic or the power-law function better 
describes the data points.


\subsection{Evolution of the Community Structure}

\begin{figure}[h]
\begin{center}
\includegraphics[width=0.3\textwidth, angle=0]{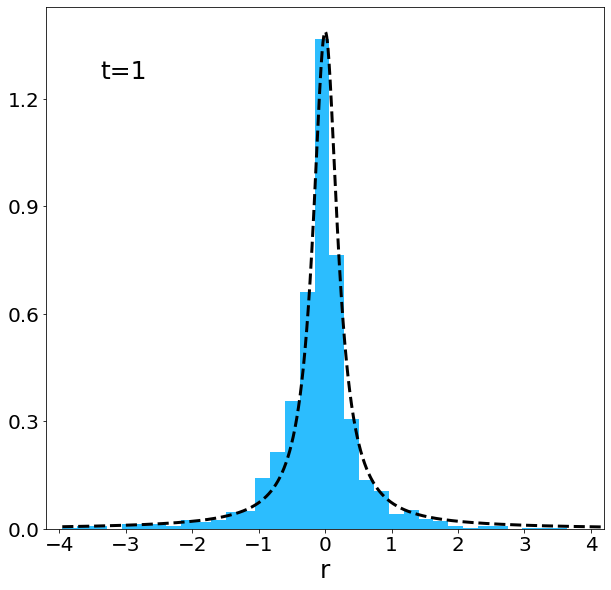}
\includegraphics[width=0.3\textwidth, angle=0]{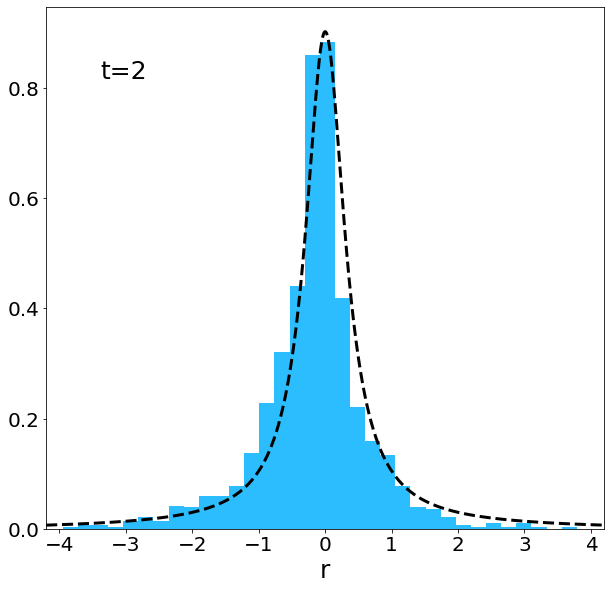}
\includegraphics[width=0.3\textwidth, angle=0]{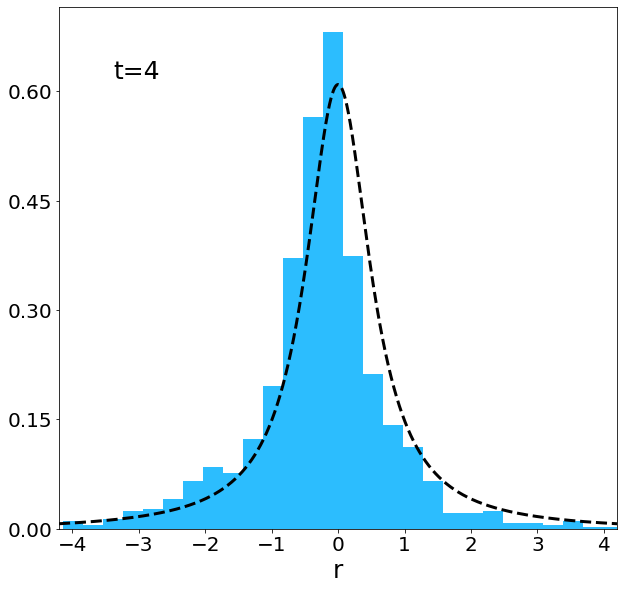}
\includegraphics[width=0.3\textwidth, angle=0]{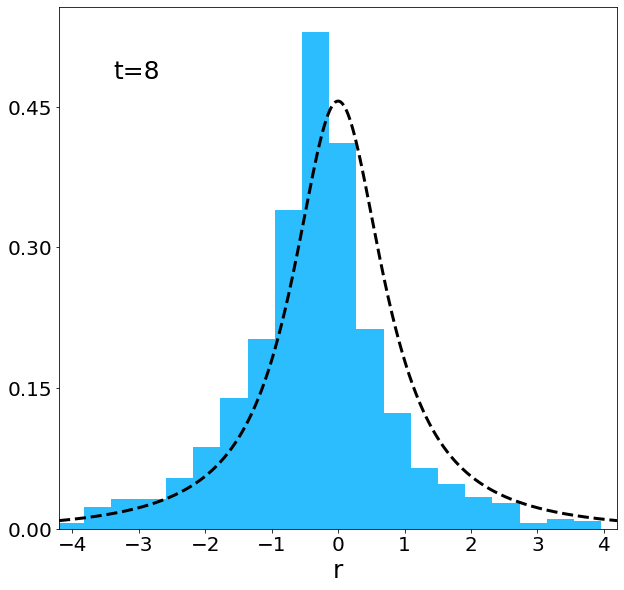}
\includegraphics[width=0.3\textwidth, angle=0]{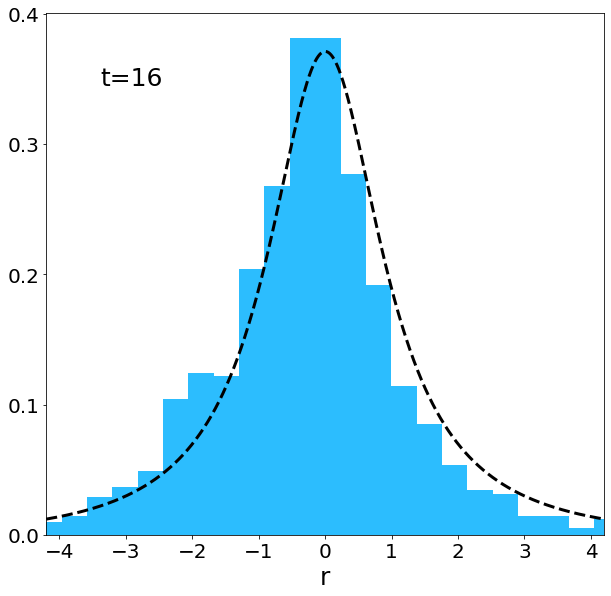}
\includegraphics[width=0.3\textwidth, angle=0]{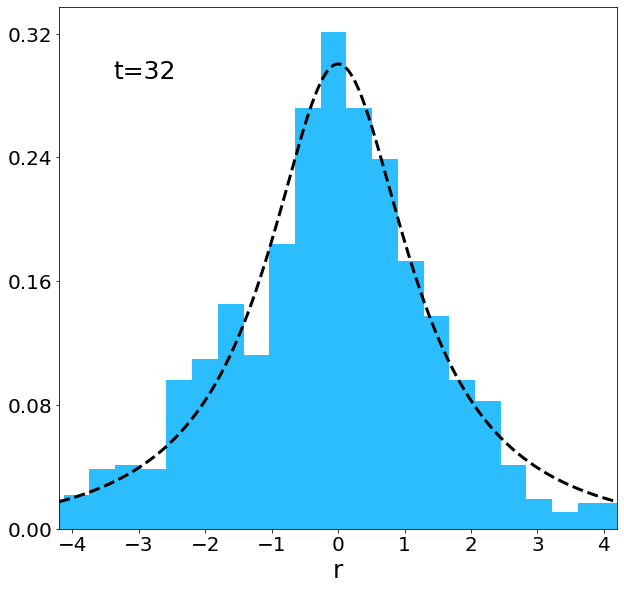}
\end{center}
\caption{\small {Species turnover distribution evaluated considering the cryptocurrencies
market share. We fix $t=1, 2, 4, 8, 16, 32$ weeks.  We consider the variable $r=log(\lambda)$
and plot the distribution $P'(r)= e^rP_{STD}(e^r,t)$, where $P_{STD}(e^r,t)$
is given by equation \ref{eq:STD}. The best fittings are represented by the dashed lines.
}
}
\label{Fig_Community2}
\end{figure}

We estimate the species turnover distribution using the cryptocurrencies
market share. We consider coins with $x(0)$ located in the stationary period.  
The fittings of the STD with the analytic expression of equation \ref{eq:STD} present contrasting results, as can be appreciated in Figure \ref{Fig_Community2}.
For some values of $t$ the fitting is satisfactory, for others  there is a small but systematic difference between fitted and empirical distributions. Empirical data present an asymmetry between left and right tails, with a greater propensity of developing negative $\lambda$ values, which correspond to decreasing populations. The parameters obtained from the fitting of the STD
can be used to produce the SAD generated by the neutral model of \cite{Pigolotti,Azaele06}.
The fitted values of B ($0.48\le B \le 0.71$) 
lead to SADs presenting a shape 
far from the expected for neutral models describing a single community without migration.\\

\begin{figure}[h]
\begin{center}
\includegraphics[width=0.95\textwidth, angle=0]{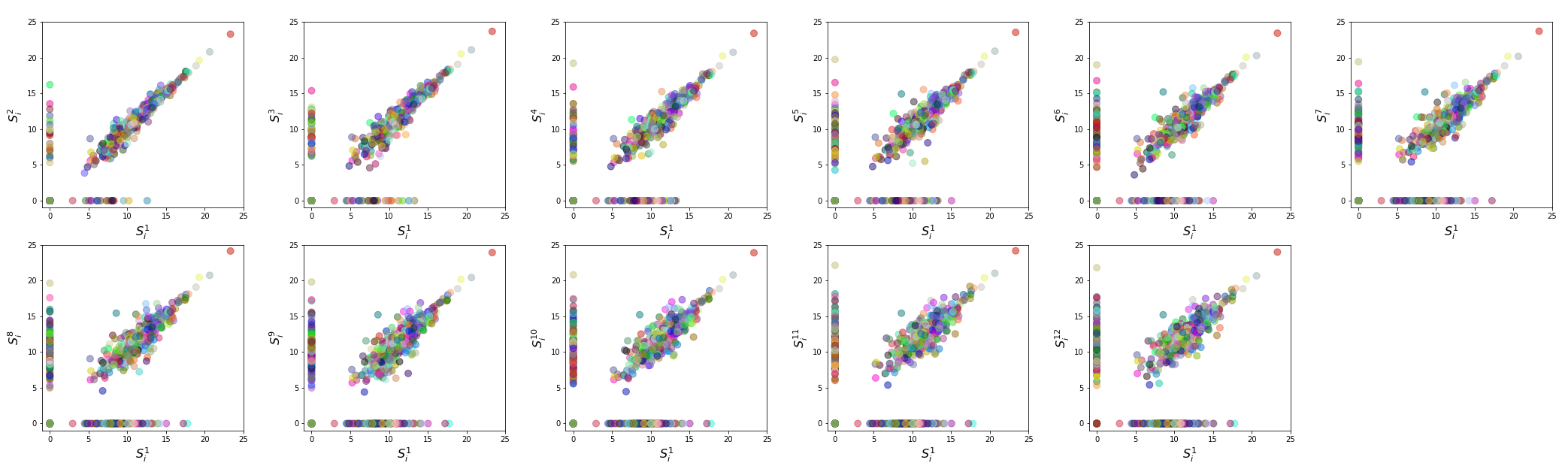}
\includegraphics[width=.95 \textwidth, angle=0]{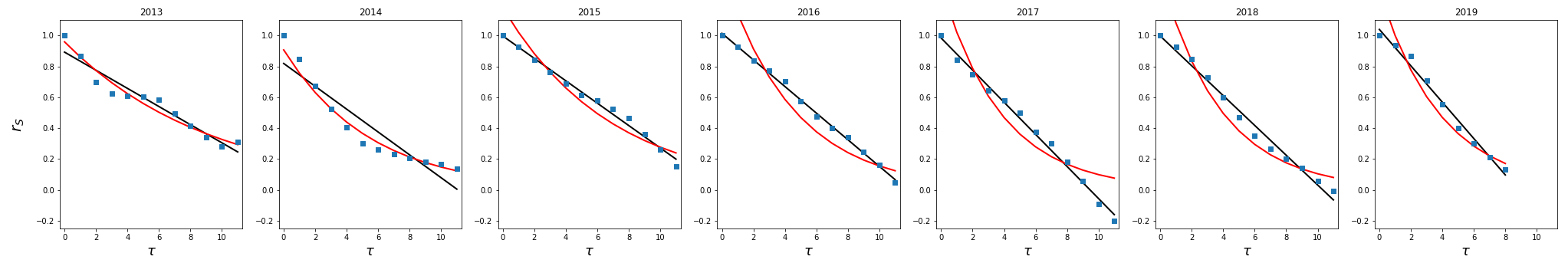}
\end{center}
\caption{\small {
{\it Top:}  Scattering plot of the log-abundances in the initial community versus abundances at the following specified time, expressed in months. Data are displayed for the interval from 2015-4-28  to 2016-4-28.
As a result, initially rare or absent species appear near the x-axis and rare or extinct species appear near the y-axis. 
 Notice how numerous cryptocurrencies go extinct. Different colors correspond to different currencies.
{\it Bottom:} The evolution of the community structure index $r_S$ during one year. Data starts from 2013-4-28.
The continuous black lines are the best fitted linear functions; red lines represent the exponential fittings.  Note that, in general, correlation presents positive values. Anyway, negative values can be reached, as a product of the final accumulation of species around the x and  y-axis.


}
}
\label{Fig_Community}
\end{figure}
\begin{figure}[h]
\begin{center}
\includegraphics[width=0.45\textwidth, angle=0]{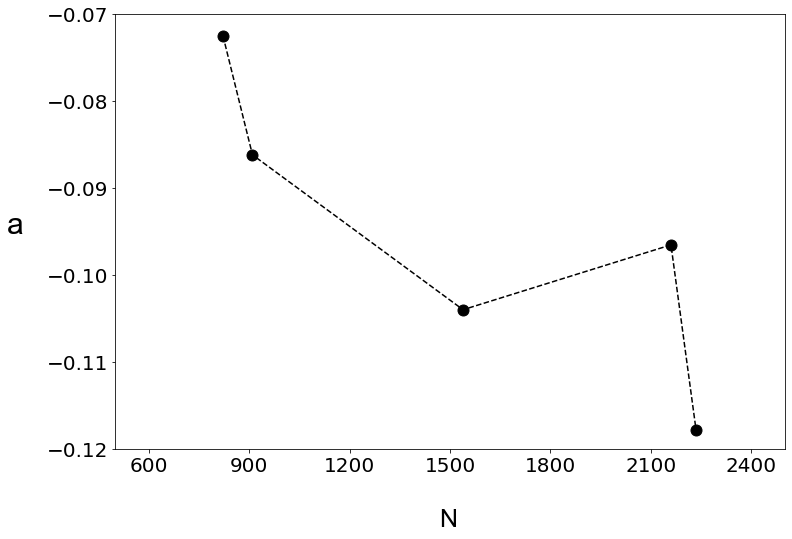}
\includegraphics[width=0.45\textwidth, angle=0]{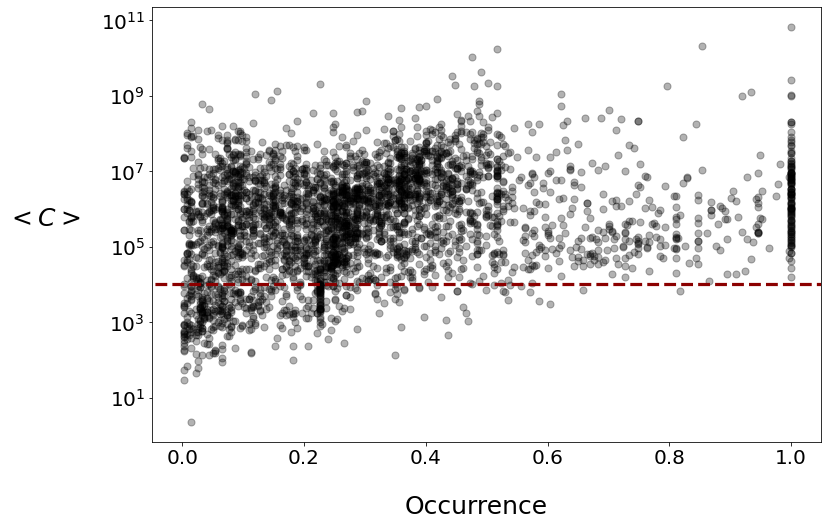}
\end{center}
\caption{\small {\it On the left:}
The regression slopes of the best fitted linear functions ($a$) versus
the number of active cryptocurrencies (species richness) for the five temporal intervals 
considered in figure \ref{Fig_Community} from 2015 to 2019. 
The first quantity is a measure of temporal $\beta$-diversity (species turnover, as the variation in species composition through time), 
the second of $\alpha$-diversity.
{\it On the right:} scattering plot of the occurrence versus the mean capitalization of a given cryptocurrency.
Data are evaluated considering the interval from the beginning of the stationary period
to the end of the dataset.
}
\label{Fig_diversity}
\end{figure}

Figure \ref{Fig_Community} shows an example of the scattering plot of the log-abundances in the initial community versus abundances at the following specified time. 
For these different plots we evaluated the correlations 
and we examined their evolution along one year.
We perform this analysis for all our data. In the period of mass radiation, we can observe an exponential behavior. In contrast, as the community enters the stationary regime, the $r_S$ decreases  with time following a clear linear behavior and this mode is maintained until the end of our time series.  
The slope values of the linear fittings ($a$) decrease along the considered years.
These 
slope values 
are an interesting parameter for quantifying what
ecologists call temporal $\beta$-diversity \cite{Magurran19}, which characterizes the change in the composition of a single community through time. 
More negative $a$ values 
represent a more intense variation in the community structure, which corresponds to a stronger temporal $\beta$-diversity.
In this framework, the relation  
between temporal $\alpha$-diversity,  which characterizes the temporal variation of 
species in a single community, and temporal $\beta$-diversity is particularly interesting. 
In Figure \ref{Fig_diversity} we can appreciate a general trend of increasing
temporal $\beta$-diversity with increasing $\alpha$-diversity. 
$\alpha$-diversity is estimated using a simple count of the number of species (species richness) in the considered interval.

In addition, we analyze if a clear relationship exists between the abundance
of the population of a given species and its lifetime. We estimated 
the abundance considering 
the mean value  of the capitalization of a species estimated over its time series. 
The lifetime is obtained by measuring the occurrence 
of the fraction of weeks 
in which a cryptocurrency is active over all the considered weeks.
By looking at the scattering plot of these quantities (Figure \ref{Fig_diversity}), 
we can see that cryptocurrencies with a capitalization smaller than $10^4$ present 
a reduced lifespan, with occurrences that hardly reach 0.5.
However, we can perceive that these quantities display an unexpected rather weak dependence.



We conclude this analysis by comparing 
the $r_S$ behavior displayed during 
the stationary period 
with the outcome of the neutral model presented in 
equation \ref{equ_model}.
Simulations are run for a number of generations that 
allows reaching $r_S$ values close to the smallest one
displayed by real data.
We fix the number of species $s$ 
considering the mean number of species appearing in the dataset
during the 
chosen periods and we select
the ratio  $2b/\sigma^2$ among values that are small enough for
generating distributions close to the Fisher's one. 
This purpose is 
reached by fixing $b=1$ and varying $\sigma$. 

\begin{figure}[h]
\begin{center}
\includegraphics[width=0.5\textwidth, angle=0]{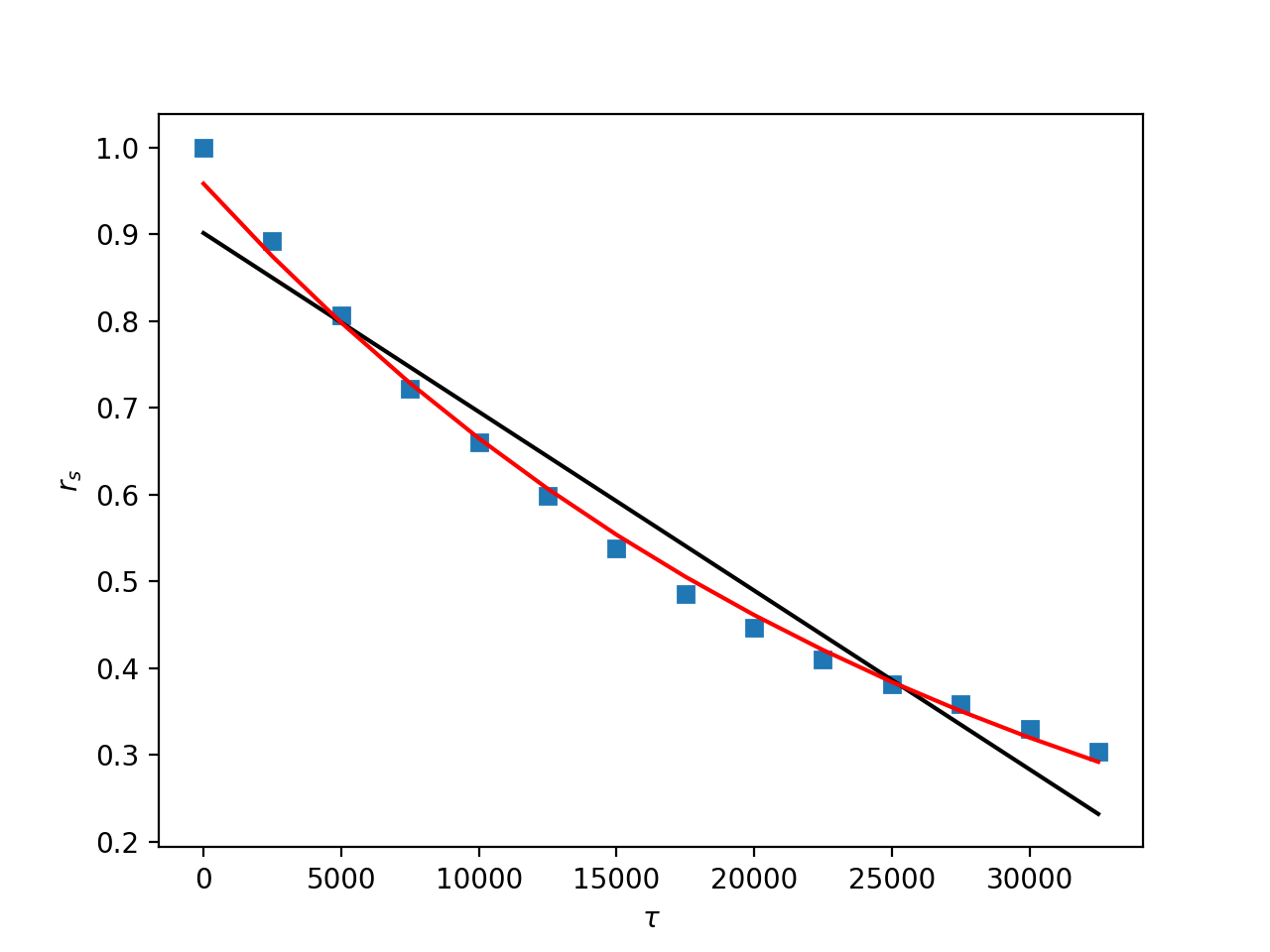}
\end{center}
\caption{\small
A typical $r_S$  decay obtained from the artificial data generated by
the neutral model of equation \ref{equ_model}.  
In this case, the model parameters are:
$N=10^7$,  
$\rho=3.5$,  
$\sigma= \sqrt{20}$, 
$b=1.$ and  
$s=1500$; $\tau$ are expressed in simulation times.
The continuous black line is the linear fitting, the red line the exponential one.
}
\label{fig_model}
\end{figure}

We collect the species abundances and take into account the extinction\slash speciation events.
These 
events are calculated 
when the term $\rho x^t_i+\sigma \sqrt{x^t_i}\eta^t_i$  crosses a zero value. 
From these artificial data, we can obtain the $r_S$ values.
We run a large amount of simulations for different values of
$N$, $\rho$ and for $\sigma> \sqrt{3}$.
For this range of parameters values, the model generates only exponential decays.
Figure \ref{fig_model} shows a typical example for specific parameters.

\subsection{Correlations}

\begin{figure}[h]
\begin{center}
\includegraphics[width=0.9\textwidth, angle=0]{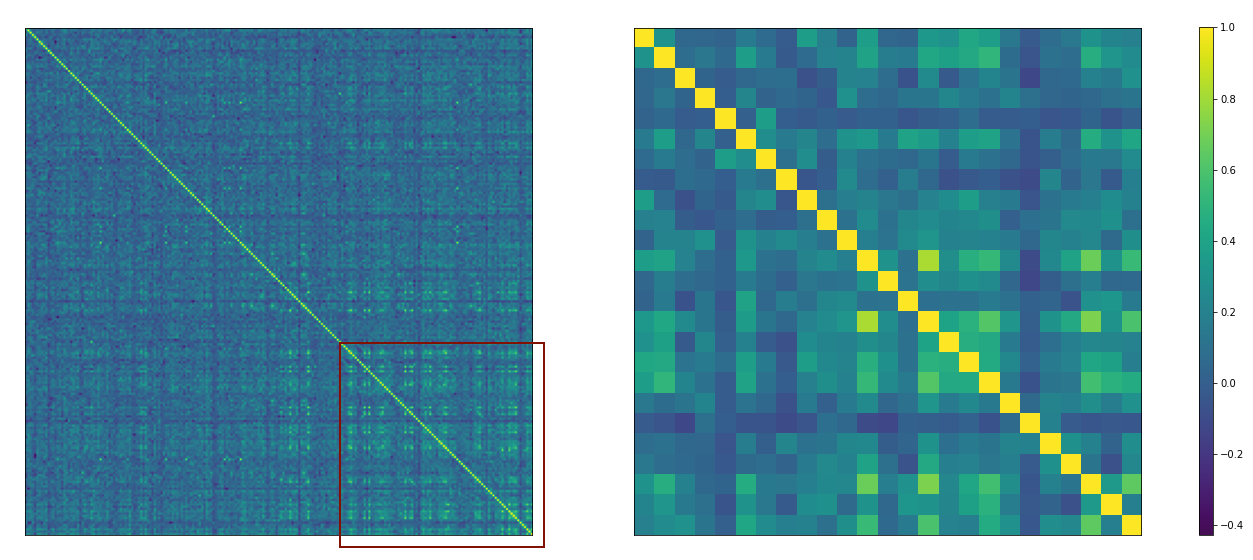}
\includegraphics[width=0.44\textwidth, angle=0]{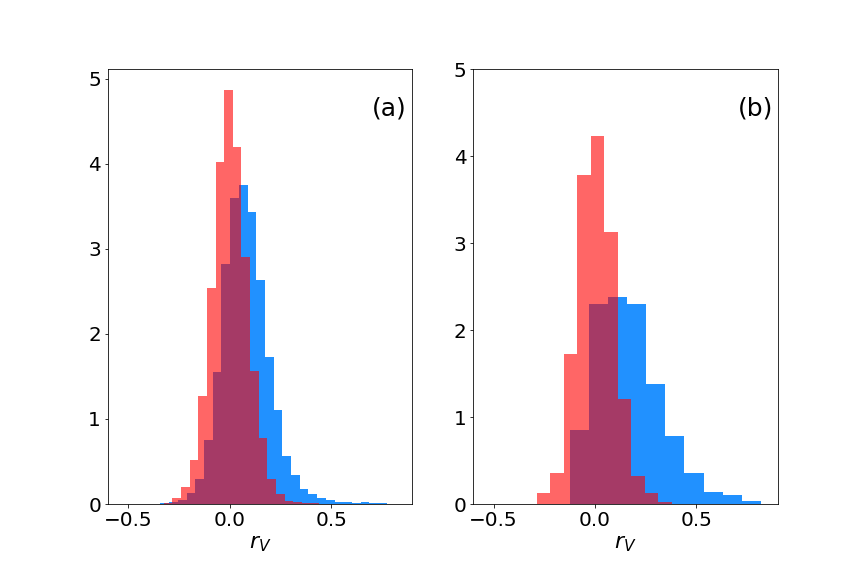}
\includegraphics[width=0.44\textwidth, angle=0]{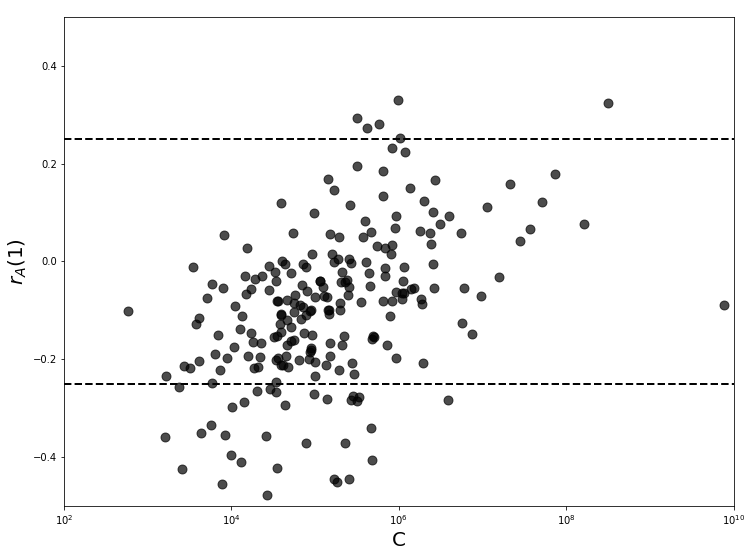}
\end{center}
\caption{\small
{\it Top:} Plots of correlation matrices for all
the considered  cryptocurrencies (left) and for the top 25 (right). 
Cryptocurrencies fill the matrix from the smallest  to the largest capitalized ones (the considered capitalization is the total capitalization over the stationary period).
Note how the pairs of highly capitalized cryptocurrencies (highlighted by the red square)
present, in general, higher correlations. 
{\it Bottom:}   On the left, in blue, are the distributions of the pair  correlations for all
the considered  cryptocurrencies (a) and for the top 25 (b).
In red, the same distributions obtained from the random shuffled series.    
On the right, the temporal  autocorrelation for $\tau=1$ 
as a function of the aggregated capitalization over the stationary period of the 
cryptocurrencies.
The dashed lines represent the 1/99 percentiles 
of the distribution of the autocorrelations of the randomized data estimated over 10 different realizations.
}
\label{Fig_Correlations}
\end{figure}

We studied the sub-set of 225 cryptocurrencies  that are 
active during the entire stationary period.
The correlation matrix for pairs of cryptocurrencies
is reported in Fig. \ref{Fig_Correlations}.
A qualitative observation of the correlation values
shows a clear deviation from the values of the randomized sample 
and the matrix presents a clear structure.
For the cryptocurrencies with higher capitalization, 
correlations have larger values, which are statistically significant and positive.
This result can be clearly appreciated by looking at the behavior of the largest 25 
cryptocurrencies.

Finally, we look at the  temporal autocorrelation 
for a single cryptocurrency: $r_A(\tau)=Corr(V(t),V(t+\tau))$. 
We focus on  $r_A(1)$, as for  $\tau>1$ autocorrelation 
values are in general indistinguishable from the ones of a random series.
As shown in Figure \ref{Fig_Correlations}, there is 
a clear dependence of $r_A(1)$ 
on the capitalization
of the considered cryptocurrencies. 
For smaller capitalizations, some cryptocurrencies present
a significant anticorrelation, for larger capitalizations only
a few cryptocurrencies show a significant, but small,
positive correlation. In particular, among the largest 25 currencies
only two present significant correlation values. 


\section{Discussion}


The shape of the SADs 
lends strong support to 
alternatives to the neutral theory. 
In fact, neutral models without migration, which describe the distribution at the 
metacommunity scale, predict a  Fisher's distribution. In contrast,
tests 
of fit quality  show that the log-normal approximates
the observed species abundance data very well.
This means that looking at the log-scale, the SAD shows a strong central tendency
and relatively few 
rare or abundant species. 
In contrast,  neutral dynamics would generate more
rare species and truly dominant ones would be improbable.


The analysis of the SPR gives no definitive answers 
that would allow the selection of one of the two alternative models.
In fact, a dependence between species and population exists,
but it is very weak and it is not possible to distinguish between a logarithmic or 
a power-law dependence.

The STD   fitting with the distribution generated by a neutral model
does not allow a straightforward interpretation.
Even if the fitting, for some values of $t$, is promising, we must
stress that the same distribution fits equally well turnover data 
generated by very different community populations, corresponding to
distinct SADs, which can  
describe situations with or without central modes.
For this reason, it is important to look at what SAD is produced by 
a given STD fitting. In our case, the STD fittings generate
SADs presenting a shape far from the expected for neutral models 
describing a single community without migration.
For this reason, we think that this result is not supporting the neutral theory.
This specific example shows how curve-fitting  is a valuable approach 
to test a model, but it is not necessarily conclusive. 
In particular, this is the case when the fitting distributions are flexible enough for 
describing very different situations. \\

The evolution of the community structure correlations shows a very interesting mode.
In fact, after the radiation phase, it displays a characteristic linear decay.
Neutral models describing a single community without migration can not generate
this linear behavior.
The absence of an exponential behavior 
suggests that the dynamics of the change in the community structure 
is very far from being random, with a quite slow drift 
in the relative abundances of moderately abundant species.
The system is apparently endowed with a mechanism, not present in the neutral model,
which allows relative rare and moderately abundant coins to persist over time. 
This conjecture is supported by the performance of the cryptocurrencies persistence
in relation to their abundance: only 
very rare coins 
effectively go systematically extinct.  
Persistence and abundance display a weak dependence
and, in contrast with the results of neutral models (see \cite{Azaele06}), 
it is not evident 
that the less abundant species are clearly more prone to extinction.


The relation of the pace of change in the community structure
with its species richness shows how the presence of more cryptocurrencies
tends to accelerate the 
reorganization of the composition of the community over time. 
If a considerable amount of new species enter the system,
the market must accommodate via a faster temporal reorganization
which can generate a finer temporal subdivision of the wealth injected in the market.
The fact that an increase in $\alpha$-diversity has a direct effect on temporal $\beta$-diversity,
may suggest that interspecific competition among cryptocurrencies is not so weak.

These original results demonstrate the particular advantage of 
using non-biological data for testing and improving
ecological theories and methodologies in the quantification
of temporal behaviors, where ecological datasets are generally 
scarce or must rely on fragmented paleontological data.\\

The analysis of the correlation matrix for pairs of cryptocurrencies
quantifies the dependence of 
the 
increase/decrease of a species $i$ on the 
increase/decrease of a species $j$.
Results clearly show the presence of a group of coins,  included among the higher capitalized 
cryptocurrencies, presenting a cohesive response in the 
variation of their capitalization.
This important result shows a coherent behavior in a sector of the market.
The positive and statistical relevant correlations suggest the presence of mutualism:
cryptocurrencies that belong to this sector benefit from an increase in the capitalization 
of the other cryptocurrencies which belong to the same sector.
This effect can be determined either by endogenous factors 
either by the 
eventual contribution of exogenous 
common cause drivers.
We can distinguish between two classes of cryptocurrencies: a minor one where 
the 
variation of the capitalization presents clear positive correlations, and a larger one, where
this variation is uncorrelated.  
At the considered time scale, there is no switch between these classes and we can label
species as belonging or not to a given class. Cryptocurrencies are not symmetric in relation to 
this behavior and the  symmetric species postulate seems not 
to be satisfied,
implying that the 
system is not neutral.
In neutral models, common species are treated simply as rare species, but 
here the behavior of the correlations suggests that different mechanisms 
are shaping these two classes of cryptocurrencies and should be taken into account.
Coins with larger capitalization can not be treated simply as 
rare coins writ large 
\cite{Gaston11}.
We remember that, in ecologically neutral models, the correlation in the abundances of a pair of species 
is the same as the correlation in the abundances of any other pair of species 
\cite{Fisher14}. Thus, a distribution with equal correlation coefficients can stand in as a proxy for measuring statistical neutrality. A similar role can be 
assumed by the correlations between pairs of $V_i(t)$.

Finally, we can shed a light on the relevance of the interactions between species,
and if they are effectively weak compared to the stochastic drivers of the dynamics, 
by contrasting the values of the correlation among pairs of cryptocurrencies with the autocorrelation
of each cryptocurrency.
The pair correlations can be seen as a proxy for between-species interdependency (interspecific) and the autocorrelations as a within-species temporal interdependency (intraspecific).
We can note how for the largest capitalizations, when the interspecific correlations are relevant and positive, the intraspecific ones are generally much smaller and statistically not significant.
In this case, interspecific correlations are not weak in comparison with intraspecific ones
and 
could be generated by some constraints on the community dynamics
not generated by neutral dynamics.\\

To sum up,  we have developed an analysis of the cryptocurrency market borrowing methods and concepts from ecology.
This approach allows for identifying specific diversity patterns and their variation, in close 
analogy with ecological systems, and to describe them effectively.
At the same time, we can contrast different ecological theories, testing the validity of
using neutral models. 
The behavior of the SAD and the evolution of the community
structure strongly suggest that these statistical patterns are not consistent with neutrality.
In particular, the necessity to increase the community composition change
when species richness increases 
suggests that the interactions among 
cryptocurrencies are not necessarily weak.
This fact is supported by the analysis of intraspecific and interspecific interdependency,
which also demonstrates that a market sector influenced by mutualistic
relations can be outlined. All these outcomes challenge
the hypothesis of weakly interacting symmetric species. 

Our results show that the community 
structure of the cryptocurrency market 
can be effectively described
by using an ecological perspective.
Our analysis, besides static distributions, highlights specific patterns 
of a rich temporal dynamics.
These data were compared directly 
with neutral models. Even if falsified, these models offer
a reference point for parsimonious description, acting as a useful null model.
Our study introduces interesting new strategies for describing
dynamical patterns that are novel even for ecological studies.
The accessibility to the whole data set of the cryptocurrency market,
without limitations derived from sampling, makes possible the 
exploration of these tools.
For these reasons, these results have an interesting impact either in the characterization of the cryptocurrency market, 
either on the relevance that 
non-biological systems can have in testing ecological theories. 
Finally, the new set of empirical regularities and general tendencies displayed by
the cryptocurrencies market could introduce interesting elements of exploration
for the exciting and emerging 
field of market ecology \cite{Farmer1,Farmer2}.

\section*{Acknowledgments}
EAAM received partial financial support from the PIBIC program of Universidade Federal do Rio de Janeiro.
We thank Prof. Marcus Vin\'\i cius Vieira, from the Graduate Program in Ecology - UFRJ, 
for useful comments and suggestions and Prof. Jorge Sim\~oes de S\'a Martins
for revising the manuscript.


\end{document}